# Effects of Reagent Rotation and Vibration on H + OH $(\upsilon, j) \rightarrow$ O + H$_2$

*Xiaohu Li,[1] Carina Arasa,[2] Marc C. van Hemert[2] and Ewine F. van Dishoeck [1, 3*]*

The dynamics of the reaction H + OH → O ($^3$P) + H$_2$ have been studied in a series of quasi-classical trajectory (QCT) calculations and transition state theory (TST) methods using high quality $^3$A' and $^3$A" potential energy surfaces (PESs). Accurate OH $(\upsilon, j)$ state resolved cross sections and rate constants on both potential energy surfaces are presented and fitted for OH at $(\upsilon = 0, j = 0 - 16)$ and $(\upsilon = 1, j = 0 - 6)$. The cross sections were calculated for different collisional energies ($E_c$), ranging from the threshold energy at each specific rovibrational state up to 1.0 eV with step sizes of 0.1 eV or less. They increase steeply with collision energy when the barrier to reaction can be overcome, after which the cross sections stay nearly constant with energy. State resolved rate constants in the temperature range 200 — 2500 K are presented based on the cross sections. Total thermal rate constants were calculated by summing the rates for reaction on the $^3$A' and $^3$A" potential energy surfaces weighted by 1/3 and taking into account the thermal populations of the rovibrational states of the OH molecules. The currently calculated thermal rate constants generally agree well with previous indirectly obtained rate constants by Tsang *et al.* (1986). It is shown that the improved canonical variational transition (CVT) treatments with the approximation of zero-curvature tunneling (ZCT) or small-curvature tunneling (SCT) produce results

*1. Leiden Observatory, Leiden University, P.O. Box 9513, 2300 RA Leiden, The Netherlands*
*2. Leiden Institute of Chemistry, Gorlaeus Laboratories, Leiden University, P.O. Box 9502, 2300 RA Leiden, The Netherlands*
*3. Max-Planck Institut für Extraterrestrische Physik, Giessenbachstrasse 1, 85748 Garching, Germany*
*Correspondence to: Prof. Dr. Ewine F. van Dishoeck; e-mail: ewine@strw.leidenuniv.nl*

more in accord with the QCT results than the TST and CVT methods. The reactions are governed by the direct reaction mechanism. The rate constants for OH in excited vibrational and rotational states are orders of magnitude larger than the thermal rate constants, which needs to be taken into account in astrochemical models.

# 1. Introduction

The O ($^3$P) + H$_2$ system has attracted interest for more than half a century.[1] In addition to its fundamental significance in chemical dynamics, it is also known to be a participant in combustion processes[2] and plays an important role in warm interstellar gas such as shocks, clouds exposed to intense UV radiation and the inner regions of protoplanetary disks.[3–5] Extensive theoretical[6–32] and experimental[1,33] investigations have been carried out to elucidate the dynamics of the O ($^3$P) + H$_2$  reaction, which has a classical energy barrier of about 0.57 eV[21] and at low temperatures occurs mainly through tunneling.[26] Thanks to long-term efforts of constructing and improving the potential energy surfaces,[6,14,17,21,25,31] classical and quantum mechanical calculations have been carried out to provide fruitful information on many aspects of this system, such as the kinetics,[34,35] isotope effects,[36] state selective dynamics,[32,37,38] intersystem crossing,[15,26] nonadiabatic effects,[7,9,13] and stereodynamics.[8,39] Among these PESs, the high quality generalized London-Eyring-Polanyi-Sato (LEPS) double-polynomial (GLDP) surfaces ($^3$A' and $^3$A") reported by Rogers et al.[21] are widely adopted in scattering simulations and can yield chemically accurate results which are in good agreement with experiments.[1,11,33] For example, the reaction cross sections measured by crossed molecular beam experiments are well reproduced by quantum wave packet calculations based on this PES.[1] Another combined experimental and theoretical investigation on the isotopic substituted reaction O ($^3$P) + $D_2$ also found good qualitative agreement between the measured dynamic results and the single-surface quasi-classical trajectory calculations.[1,33] These studies also showed that intersystem crossing with the excited singlet states have at most a very minor effect on the reaction probabilities. In fact, according to the exact quantum study of spin-orbit-induced intersystem crossing by Chu et al.,[15,40] the effects of spin-orbit coupling on the total reaction cross sections are insignificant if the fine-structure resolved cross sections in the O ($^3P_{2,1,0}$) + $H_2$ reaction are treated statistically. The quantum scattering calculations performed by Balakrishnan et al.[13,18,20] mainly reported thermal rate coefficients, and found the reaction probability of the $^3$A' PES to be slightly smaller than that of the $^3$A" PES in the energy range of 0.5 — 1.2 eV. Brandão et al.[17] extended the $^3$A" PES from Rogers et al.[21] with van der Waals interactions and used it to study the influence of the long-range potential on the reaction rates, especially at (ultra) cold temperatures, where tunneling and resonances are significant.

There are far fewer studies of the reverse reaction. The most detailed calculations have focused on the cross sections, thermal rate constants and rate coefficients of the isotope exchange reaction D + OH → OD + H, which plays an important role in interstellar clouds.[41–44] However, few studies have been done for the more basic process

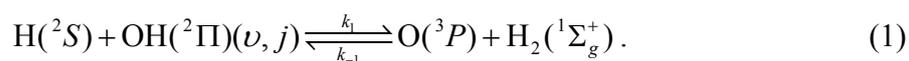

$$\text{H}(^2S) + \text{OH}(^2\Pi)(\upsilon, j) \underset{k_{-1}}{\overset{k_1}{\rightleftharpoons}} \text{O}(^3P) + \text{H}_2(^1\Sigma_g^+)\,. \qquad (1)$$

This is partly because obtaining a beam of OH is experimentally challenging.[45] The rate constant for $k_1$ in Eq. 1 offered in the National Institute of Standards and Technology (NIST) Chemical Kinetics Database,[46] which is often employed in astrochemical models but dates from 1986, was derived from the rate constant of the reverse reaction,

$$k_1 = K / k_{-1} \qquad (2)$$

where $K$ is the kinetic reactive rate constant. The $k_{-1}$ was suggested according to a combined experimental and TST investigations where the transition state theory (TST) calculations were based on an older PES.[47] Since the uncertainty of $k_1$ arises from the uncertainties in both $K$ and $k_{-1}$, determination of more reliable rate constants for this reaction using a more accurate PES are warranted. Moreover, there is a growing need in astrochemical models for state-specific rate constants in addition to thermal rates.[48–51] Specifically, non-thermal excitation of OH in high rotational and/or vibrational states has been observed in shocks[48] and in the inner regions of protoplanetary disks,[49–51] which may drive the H + OH reaction. In these regions, gas temperatures are up to 2000 K.

In this paper, we have carried out the first QCT, TST, and variational transition state theory (VTST) calculations for the title reactions on the $^3A'$ and $^3A''$ GLDP PESs of Rogers *et al.*[21] OH ($v$, $j$) resolved rate constants and thermal rate constants are reported. Other than using the approximate ZCT and SCT methods, tunneling is ignored in the paper due to the classical nature of the QCT approach. The paper is organized as follows: in Section 2 we briefly review the theoretical methodologies. In Section 3, results and discussion are presented. Finally, Section 4 closes with the conclusions.

## 2 Methods

### 2.1    Potential Energy surface

Scattering calculations were performed on the $^3A'$ and $^3A''$ potential energy surfaces of Rogers *et al*,[21] which were until recently the most accurate published surfaces for this reaction. These surfaces, generated from 951 geometries and with high-accuracy for a subset of 112 of these geometries, were based on complete-active-space self-consistent-field (CASSCF) computations using correlation-consistent basis sets. Two different fitting methods, the rotating Morse spline (RMOS) method and the GLDP method, were employed and compared to each other and also to a previous empirical LEPS surface. Overall, the GLDP fit and RMOS fit to the *ab initio* data are very similar, but the GLDP fit has a higher fitting accuracy and should yield accurate reactive scattering calculations. The GLDP PESs employed in our work consist of two

terms:

$$V_{GLDP} = V_{GLEPS} + V_{DPOLY} \tag{3}$$

$V_{GLEPS}$ is a Generalized LEPS term and $V_{DPOLY}$ a sum of two high-order polynomials (as functions of the three interatomic distances) multiplied by switching functions. In $V_{GLEPS}$, a cubic spline fit of near-asymptotic *ab initio* data for each of the three isolated diatomic molecules is used instead of the Morse function of the usual LEPS function. Further details can be found in the original reference.[21]

A typical A+BC reaction can lead to four possible scattering channels: A+BC, AB+C, AC+B, and A+B+C. We have focused on the H + OH → O + H$_2$ abstraction reaction both on the $^3$A' and on the $^3$A'' PESs. The two potential energy surfaces have similar characteristics and are degenerate for collinear configurations. Both of them are dominated by a barrier and have a broad van der Waals well along the entrance valley. The barrier heights on the $^3$A' and $^3$A'' PESs are 0.476 and 0.468 eV, and the depths of the wells are 0.009 and 0.004 eV, respectively. The differences in barrier heights and van der Waals depths in the two PESs are smaller than the root-mean-square (rms) error ( ±0.0129 eV) induced by the fitting procedure and are in any case not relevant for the temperature range considered here. However, these small differences may be responsible for slightly different reaction thresholds, and also play a role at in quantum mechanical studies at low collision energies and stereodynamics studies.

The corresponding energy-level diagram for different rovibrational states $(v, j)$ of the OH molecule is shown in Figure 1. The rovibrational energy levels $E_{v,j}$ were calculated from the PESs using the QCT code and are presented both with and without the zero-point energies. In practice, the zero-point energy of OH molecule needs to be taken into account for the reactive channel. In our case, $j$ is the purely rotational angular momentum of the OH molecule (often denoted as $N$ in the spectroscopic literature) since in our work electronic orbital and spin angular momenta are discarded. Each OH rotational level $j$ modeled here, is in reality split into two different spin states, $^2\Pi_{1/2}$ and $^2\Pi_{3/2}$, which have an energy difference of only 0.017 eV.[52] This difference is ignored in our calculations.

## 2.2    Quasi-classical trajectory calculations

In this work, we have used the Venus96 QCT code developed by Hase *et al.*[53] to compute cross sections as functions of energy, from which state-resolved rate constants as well as the thermal rate constants are subsequently calculated. In all cases, the initial parameters are as follows. An initial distance of 6.4 Å is adopted between the attacking H atom and the center of mass of the target OH molecule. Batches of $2\times10^4$ trajectories have been carried out for each collision energy of the incident hydrogen atom. The integration step size is chosen to be 0.1 fs, which is small enough to ensure the conservation of the total energy and the total angular momentum. To calculate thermal rate constants, for both $^3$A' and $^3$A" PESs, the initial rovibrational conditions of the OH molecule were chosen to include the range of states $(\upsilon=0, j=0-16)$ and $(\upsilon=1, j=0-6)$. To evaluate the cross sections, the collision energies ($E_c$) were set up on a finely spaced grid: on the $^3$A' PES, for $\upsilon = 0$ at $j = 0$ up to 16, $E_c$ were chosen as follows: from 0.2 to 0.32 with a step size of 0.02 eV; 0.35 eV, and from 0.4 to 1.0 eV with a step size of 0.1 eV. For $\upsilon = 1$ at $j = 0$ up to 6, $E_c$ were chosen at 0.06, 0.08, 0.10, 0.12, 0.15, 0.2, 0.25, and from 0.3 to 1.0 eV with a step size of 0.1 eV. For the $^3$A" state, the selected $E_c$ are the same as those on $^3$A' state except those at $\upsilon = 0$ and $j = 7$ up to 16, which employed the same energy values as for $\upsilon = 1$, $j = 0$ up to 6 on the $^3$A' PES.

During the QCT simulations, the impact parameter $b$ is sampled from $0 \leq b \leq b_{max}$ by $b = b_{max} \, \xi_0^{1/2}$, where $\xi_0$ is a pseudorandom number, and the maximum impact parameter, $b_{max}$, is the value of $b$ for which the reactive probability drops to zero when $b \geq b_{max}$. In our work, $b_{max}$ is optimized by trial calculations with a range of $b_{max}$ values at fixed initial condition of the reactants, until the reactive probability of the observed channel is zero while $b = b_{max}$. The criterion is that at most one reactive trajectory occurs among a large sum of total trajectories (for example, $10^4$), thus the reaction probability is approximately zero. Generally, 3 — 6 trials are necessary to optimize $b_{max}$ at a fixed $E_c$. In the present calculations, $b_{max}$ has been determined by running batches of $10^4$ trajectories. Overall, the value of $b_{max}$ increases with increasing rovibrational state of the OH reactant at each fixed $E_c$, leading to a corresponding enhancement of the cross section as well as the relevant rate constant. According to our experience, a slightly higher $b_{max}$ does not affect the cross sections much, whereas a too small value will result in significant errors.

The integral cross section, $\sigma(\upsilon, j)$, and the associated uncertainties $\Delta\sigma(\upsilon, j)$ at a certain rovibrational level $(\upsilon, j)$ can be determined from[54]

$$\sigma(\upsilon, j) = \pi b_{max}^2 \frac{N_r(\upsilon, j)}{N(\upsilon, j)}, \quad \Delta\sigma(\upsilon, j) = \sqrt{\frac{N(\upsilon, j) - N_r(\upsilon, j)}{N(\upsilon, j) N_r(\upsilon, j)}} \sigma(\upsilon, j) \qquad (4)$$

where $N$ and $N_r$ denote the total number of sampled and reactive trajectories, respectively. Eq. (4) is only valid when the $\xi_0^{1/2}$ sampling method is used. Since we have employed a large number of trajectories, the standard error is $\pm 20\%$ for collisional energies near threshold and less than $\pm 6\%$ when the collisional energy is more than 0.4 eV.

Subsequently, the specific rate constant at a temperature $T$ for an initial rovibrational $(\upsilon, j)$ state of the OH molecule is calculated from the integral cross sections by averaging over a Maxwell distribution of energies through

$$k(T; \upsilon, j) = \sqrt{\frac{8k_B T}{\pi \mu_R}} (k_B T)^{-2} \int_0^{1.0} E_c \sigma(\upsilon, j) \exp(\frac{-E_c}{k_B T}) \, dE_c \qquad (5)$$

where $k_B$ is the Boltzmann constant, $\mu_R$ the reduced mass of the reactants, and $E_c$ the collision energy.

The thermal rate constant for an initial vibrational state can be written as

$$k(T) = \sum_{\upsilon=0} \sum_{j=0} p_{\upsilon j}(T) \, k(T; \upsilon, j) \qquad (6)$$

where $p_{\upsilon j}(T)$ are the fractional population of the OH molecule for a thermal distribution at temperature $T$, calculated by

$$p_{\upsilon,j}(T) = (2j+1)\exp(-E_{\upsilon,j}/k_BT)/Q_{\upsilon,j}(T) \qquad (7)$$

with

$$Q_j(T) = \sum_{\upsilon}^{\infty} \sum_{j=0}^{\infty} (2j+1)\exp(-E_{\upsilon,j}/k_BT) \qquad (8)$$

The fractional populations of the OH molecule as functions of $j$ at various temperatures are shown in Figure 2. This figure demonstrates that $j$ levels up to 16 in $\upsilon = 0$ indeed need to be included at higher temperatures. In the calculations, the sum of Eq. (6) is truncated at $j = 16$ for $\upsilon = 0$ and $j = 6$ for $\upsilon = 1$ for both PESs. The rate constants have been calculated by summing the rates on each of the potential energy surfaces ($^3$A' and $^3$A") weighted by 1/3.[27–30] This is because the overall reaction leads to 3 triplet potential energy surfaces, 2 having A" symmetry, and 1 A'. One of the A" symmetry PESs does not contribute to the reaction since it has a higher energy barrier.[12] Finally, to make these data available to astronomers in a convenient form, the specific rate constants for individual OH $(\upsilon, j)$ states were fitted to the Arrhenius form, $k(T) = \alpha \times (T/300)^{\beta} \times e^{(-\gamma/T)}$, where $T$ is the gas temperature, $\alpha$ is known as the pre-exponential factor, $\beta$ dictates the temperature dependence of the rate coefficient and $\gamma$ is the activation energy of the reaction in units of K.

## 2.3 Transition state theory

Transition state theory is a useful tool for calculating approximate rate constants for reactions occurring in the gas phase and in the condensed phases based on the barrier height, the density of states at the transition state $Q$, and on the (products of the) partition functions of the reactants $\Phi$. The rate constant calculations can be improved by the variational transition state theory (VTST), where a dividing surface is defined so that all reactive trajectories must pass through it. The dividing surface is defined along the reaction coordinate $s$, which is the distance along the minimum energy path (MEP), with $s = 0$ at the saddle point, negative on the reactant side, and positive on the product side. The rate constants according to the generalized transition state theory (GTST) as a function of temperature $T$ and $s$ is given by[55,56]

$$k^{GTST}(T,s) = \frac{\alpha}{\beta h}\frac{Q^{GTST}(T,s)}{\Phi^R(T)}e^{-\beta V_{MEP}(s)} \qquad (9)$$

where $\alpha$ is the number of symmetry equivalent reaction paths, $\beta = \dfrac{1}{k_B T}$, and $h$ is Planck's constant. $Q^{GTST}(T,s)$ and $\Phi^R(T)$ are the density of states at the generalized transition state at point $s$ and the partition function of the reactants, respectively. $V_{MEP}(s)$ is the potential energy at point $s$ along the minimum energy path. The VTST rate constant $k^{VTST}$ is the minimum value of the GTST rate constant (Eq. 9) at a given $T$:

$$k^{VTST}(T) = \min_s k^{GTST}(T,s) = k^{GTST}(T,s^{CVT}(T)) \qquad (10)$$

where $s^{CVT}(T)$ is the location of the Canonical Variational Transition (CVT) state at a given temperature $T$. Quantum effects may be important for light systems and at lower temperatures, especially for tunneling along the reaction-coordinate motion. In order to include these effects, the $k^{VTST}(T)$ is multiplied by a transmission coefficient, usually by the approximation of Zero Curvature Tunneling (ZCT) or Small Curvature Tunneling (SCT).[55] Here, TST, CVT, CVT/ZCT and CVT/SCT have been employed to calculate the rate constants on the $^3$A' and $^3$A" surfaces for the $(\upsilon = 0, j = 0)$ state of the reactant molecule, yielding the corresponding rate constants in the temperature range of 200 — 2500 K. In all transition state calculations the polyrate program[57] was used.

## 3. Results and Discussion

### 3.1 Integral cross sections

Cross sections starting from various initial excitation states, $(\upsilon, j)$, of the OH molecule have been computed to obtain the thermal rate constant for the title reaction. For $^3$A' and $^3$A" PESs, integral cross sections for $(\upsilon = 0, j = 0 - 16)$ and $(\upsilon = 1, j = 0 - 6)$ have been calculated. All of these cross sections are fitted to the formula $\sigma = P_1 \times (1 - (P_2 / E_c)^2)^3 \times E_c^{P_3}$ and listed in Table I, here $P_2$ gives the threshold energy of the corresponding reaction.

In Figure 3, the integral cross sections as functions of collision energies $E_c$ (the excitation functions) are displayed for selected $(\upsilon, j)$ states. The two PESs show a similar behavior which is not surprising since they are very comparable, see section 2.1. For both PESs, the cross sections increase with increasing $E_c$, which is common for barrier-dominated reactions. Basically, the rovibrational excitation of the OH molecule increases the cross sections for $(\upsilon = 0, j)$ states. For states with $(\upsilon = 1)$, the reagent rotational excitation does not always help the reaction. In particular, the cross sections at lower collision energies (< 0.5 eV) decrease with increasing rotational excitation of OH molecule. This is probably because the increasing rotational quantum number $j$ induces a barrier in the entrance valley of the potential energy surfaces, which will be especially relevant at low collision energies. Note that the errors in our QCT results near the threshold energy are large due to the very low reaction probabilities. In addition, the QCT method does not include tunneling which could be important near threshold.

It is seen that the cross sections for different $(\upsilon, j)$ states are larger for the $^3A''$ PES than for the $^3A'$ surface, indicating that the $^3A''$ surface plays a more important role in the rate constants. The same conclusion was found previously in quantum mechanical investigations of the reverse reaction, O + $H_2$ in the energy range 0.5 — 1.2 eV.[20]

We "observed" a large number of trajectories over the two PESs in every rovibrational state and found that all reactions are dominated by the direct reaction mechanism, that is, the attacking atom $H_a$ collides with the $OH_b$ molecule and forms an $H_2$ that leaves immediately, indicating that the broad van der Waals well is too shallow to "trap" any atom, even near the threshold collision energy. Two typical trajectories are shown in Figure 4.

## 3.2 Rate constants

The specific reaction rates for various OH $(\upsilon, j)$ states as well as thermal rate constants were derived from the fitted cross sections, and then fitted to the Arrhenius form. By fitting the data separately to two temperature ranges (200 — 1000 K and 1000 — 2500 K) on a logarithmic scale, the fitting errors were significantly reduced compared to the fitting throughout the overall temperature range. The fitted parameters are listed in Table II.

Selected rate constants for various OH $(\upsilon, j)$ states as a function of inverse temperature are shown in Figure 5. As expected, the rate constants increase with increasing temperature and follow an Arrhenius trend. For both PESs, the vibrational excitation of OH evidently accelerates the reaction. The rotational excitation of OH accelerates the reaction for $(\upsilon = 0, j)$ states except for $j = 0$ on the ³A' PES. For $(\upsilon = 1, j)$ states, similar to the conclusions from the corresponding cross sections, rotational excitation reduces the rate constants due to the induced effective barriers at lower temperatures.

Figure 6 compares the calculated rate constants for the $(\upsilon = 0, j = 0)$ level summed over both PESs by the different QCT, TST, CVT, CVT/ZCT, and CVT/SCT methods in the temperature range of 200 — 2500 K. The calculated TST and CVT rate constants are smaller than the QCT results in the temperature range below 500 K, whereas the improved CVT treatments, namely zero-curvature tunneling (ZCT) and small-curvature tunneling (SCT) methods, which attempt to take into account tunneling effects, indeed produce more accurate rate constants for the title reaction.

The outcome of the transition state calculations is rather surprising. At low temperatures the TST rate is four orders of magnitude smaller than the QCT rate, but both rates are, due to the ~ 0.5 eV barrier, very low anyway. Both methods could at these low temperatures suffer from neglect of tunneling. The standard CVT method produces the same rates as pure TST at the low temperatures, but adding corrections for tunneling according to the SCT formalism makes the CVT/SCT rates agree with the QCT rates. This agreement between the QCT and CVT/SCT results at low temperatures must be largely fortuitous since QCT does not take tunneling into account. The question of the importance of tunneling can only be addressed properly by a full quantum mechanical (QM) calculation of the OH + H reaction rates on the triplet potential energy surfaces. Such calculations have not yet been performed and thus one can only speculate on the basis of results for other atom-diatom exchange reactions with a barrier for which full QM calculations have been performed. In the text book example of such a reaction, the H + $D_2$ hydrogen/deuterium exchange reaction, as discussed e.g. in Ref. 55, there is near perfect agreement (within 5% on average) between full QM and QCT rates even at temperatures as low as 200 K, suggesting that for the H + $D_2$ case tunneling is less important. In that study, also a CVT variant which includes tunneling corrections, named MCPVAG, was found to produce the same rates as were obtained with QCT and QM, while the native TST and CVT gave nearly identical but much lower rates. We therefore conclude that good agreement between QCT and CVT/SCT rates can be reached as found in both our studies and that in Ref. 55, but for reasons that are not yet understood.

Figure 7 shows a blow-up of the QCT results in the 200-1000 K range on a linear scale. This figure also includes the previous studies by Tsang et al.[58] whose rate constants were derived from the reverse O + $H_2$ reaction (over the 298 — 2500 K range).[30] The 1986 Tsang et al.'s results are in good agreement with the present thermal QCT results, which were calculated using a more accurate benchmark PESs, with differences typically 50%.

Another interesting phenomenon in Figure 7 is that the values of the thermal rate constants are higher than those at $(\upsilon = 0, j = 0)$ state when $T > 600$ K, but lower when $T \sim 200$ K. This is because at higher temperatures, the higher rotational excitation states, $(\upsilon = 0, j = 3-10)$, contribute more to the thermal rate constants, as seen in Figure 2. The values of the cross sections at those higher $j$ states are significantly higher, especially at higher energies, see Figure 3. Therefore, the thermal rates are also higher.

The importance of the state-selective rate constants computed here for astrochemical applications is demonstrated with two examples. In interstellar shocks with kinetic temperatures $T_{kin}$ of 500-2000 K, the OH molecules lose their internal energy more rapidly than their translational energy, so the rotational temperature $T_{rot}$ characterizing the OH excitation is much lower than $T_{kin}$. In this case, the state-specific rate constants for $j=0$ will be lower than the thermal rate constants at high temperatures. The difference is 30% at 1000 K, increasing to 67% at 2000 K. A reverse situation is provided by the warm (few hundred K) surface layers of protoplanetary disks where OH may be produced by Lyman $\alpha$ photodissociation of $H_2O$ in either vibrationally excited and/or highly excited rotational levels with values observed up to $j = 35$.[49–51] In this case the appropriate rate constants for the OH $(0, j)$ + H reaction are orders of magnitude larger than the thermal ones, as shown in Figure 7.

## 4. Conclusions

Using accurate *ab initio* PESs, integral cross sections and state-resolved rate constants were calculated for various OH $(\upsilon, j)$ states for the reaction OH + H $\rightarrow$ O + $H_2$ and presented in fitted form for temperatures > 200 K. Moreover, thermal rate constants are reported. The reactions occurring on the $^3$A' and $^3$A" PESs are dominated by the direct reaction mechanism, belonging to the abstraction type. The $^3$A" PES plays a more important role in the rate constants at lower temperatures whereas the $^3$A' surface contributes at higher temperatures. Both the rotational and vibrational excitation of the OH molecules will enhance the reaction rates at higher collision energies, but the vibrational excitation has a larger impact than the rotational excitation. In particular, an effective barrier may be induced by the rotational excitation at lower collision energies (< 0.6 eV).

The rate constants presented here should be more accurate than previous values used in astrochemical models of warm interstellar gas. The state-selective values are found to make orders of magnitude differences compared with thermal ones.

**Table** I Parameters of the fitting formula $\sigma = P_1 \times (1-(P_2/E_c)^2)^3 \times E_c^{P_3}$ for the cross sections $\sigma$ (Å$^2$) as a function of collision energy $E_c$ for various OH $(\upsilon, j)$ states on the $^3$A' and $^3$A'' potential energy surfaces.

| $(\upsilon, j)$ | $^3$A' | | | $^3$A'' | | |
| --- | --- | --- | --- | --- | --- | --- |
| | $P_1$ | $P_2$ | $P_3$ | $P_1$ | $P_2$ | $P_3$ |
| (0, 0) | 0.328 | 0.215 | 0.226 | 0.456 | 0.235 | 0.730 |
| (0, 1) | 0.358 | 0.217 | 0.420 | 0.446 | 0.237 | 0.625 |
| (0, 2) | 0.355 | 0.218 | 0.610 | 0.441 | 0.234 | 0.637 |
| (0, 3) | 0.359 | 0.225 | 0.572 | 0.431 | 0.234 | 0.636 |
| (0, 4) | 0.380 | 0.235 | 0.602 | 0.437 | 0.232 | 0.516 |
| (0, 5) | 0.402 | 0.238 | 0.638 | 0.460 | 0.224 | 0.631 |
| (0, 6) | 0.407 | 0.236 | 0.724 | 0.455 | 0.212 | 0.620 |
| (0, 7) | 0.426 | 0.220 | 0.890 | 0.463 | 0.135 | 1.181 |
| (0, 8) | 0.414 | 0.217 | 0.676 | 0.476 | 0.132 | 1.010 |
| (0, 9) | 0.440 | 0.203 | 0.745 | 0.525 | 0.129 | 0.971 |
| (0, 10) | 0.482 | 0.206 | 0.620 | 0.592 | 0.129 | 0.876 |
| (0, 11) | 0.513 | 0.201 | 0.590 | 0.633 | 0.125 | 0.772 |
| (0, 12) | 0.555 | 0.196 | 0.573 | 0.715 | 0.123 | 0.697 |
| (0, 13) | 0.613 | 0.182 | 0.701 | 0.715 | 0.123 | 0.697 |
| (0, 14) | 0.661 | 0.180 | 0.630 | 0.840 | 0.116 | 0.473 |
| (0, 15) | 0.719 | 0.183 | 0.532 | 0.898 | 0.108 | 0.377 |
| (0, 16) | 0.747 | 0.181 | 0.407 | 0.959 | 0.094 | 0.391 |
| (1, 0) | 0.794 | 0.018 | 0.172 | 1.101 | 0.025 | 0.317 |
| (1, 1) | 0.841 | 0.034 | 0.212 | 1.124 | 0.026 | 0.429 |
| (1, 2) | 0.859 | 0.052 | 0.257 | 1.042 | 0.058 | 0.239 |
| (1, 3) | 0.845 | 0.071 | 0.211 | 1.017 | 0.061 | 0.254 |
| (1, 4) | 0.894 | 0.074 | 0.333 | 1.056 | 0.062 | 0.329 |
| (1, 5) | 0.960 | 0.076 | 0.469 | 1.077 | 0.062 | 0.355 |
| (1, 6) | 0.988 | 0.076 | 0.547 | 1.106 | 0.061 | 0.341 |

**Table** II Thermal rate constants fitted to $k(T) = \alpha \times (T/300)^{\beta} \times e^{(-\gamma/T)}$ (cm$^3$ molecule$^{-1}$ s$^{-1}$).

| $(\upsilon, j)$ | 200 — 1000 K | | | 1000 — 2500 K | | |
| --- | --- | --- | --- | --- | --- | --- |
| | $\alpha$ | $\beta$ | $\gamma$ | $\alpha$ | $\beta$ | $\gamma$ |
| (0, 0) | 2.32E-16 | 1.448 | 2597 | 1.60E-11 | 0.031 | 3992 |
| (0, 1) | 1.05E-16 | 1.557 | 2621 | 1.93E-11 | 0.012 | 4114 |
| (0, 2) | 1.25E-16 | 1.527 | 2656 | 1.60E-11 | 0.030 | 4124 |
| (0, 3) | 2.42E-16 | 1.440 | 2762 | 1.83E-11 | 0.013 | 4182 |
| (0, 4) | 4.37E-16 | 1.374 | 2881 | 2.00E-11 | 0.010 | 4227 |
| (0, 5) | 1.12E-16 | 1.553 | 2723 | 2.09E-11 | 0.008 | 4230 |
| (0, 6) | 8.32E-17 | 1.584 | 2544 | 1.35E-11 | 0.059 | 4043 |
| (0, 7) | 3.76E-17 | 1.669 | 2105 | 4.04E-12 | 0.200 | 3582 |
| (0, 8) | 8.20E-17 | 1.579 | 2047 | 5.03E-12 | 0.179 | 3439 |
| (0, 9) | 6.30E-17 | 1.614 | 1814 | 3.01E-12 | 0.247 | 3186 |
| (0, 10) | 1.67E-16 | 1.514 | 1914 | 5.34E-12 | 0.197 | 3225 |
| (0, 11) | 1.44E-16 | 1.535 | 1679 | 3.70E-12 | 0.248 | 2973 |
| (0, 12) | 2.45E-16 | 1.481 | 1626 | 3.96E-12 | 0.253 | 2866 |
| (0, 13) | 3.31E-16 | 1.447 | 1647 | 3.87E-12 | 0.260 | 2845 |
| (0, 14) | 9.26E-16 | 1.330 | 1431 | 3.61E-12 | 0.285 | 2509 |
| (0, 15) | 1.93E-15 | 1.245 | 1328 | 3.67E-12 | 0.294 | 2333 |
| (0, 16) | 2.16E-15 | 1.233 | 1148 | 3.01E-12 | 0.324 | 2127 |
| (1, 0) | 1.49E-13 | 0.722 | 307 | 3.90E-12 | 0.312 | 751 |
| (1, 1) | 1.69E-13 | 0.711 | 531 | 4.82E-12 | 0.290 | 991 |
| (1, 2) | 1.44E-13 | 0.735 | 750 | 6.21E-12 | 0.262 | 1261 |
| (1, 3) | 1.55E-13 | 0.735 | 979 | 1.14E-11 | 0.192 | 1543 |
| (1, 4) | 8.03E-14 | 0.828 | 1123 | 1.43E-11 | 0.170 | 1782 |
| (1, 5) | 2.59E-14 | 0.973 | 1064 | 1.07E-11 | 0.207 | 1816 |
| (1, 6) | 1.02E-14 | 1.085 | 883 | 6.33E-12 | 0.268 | 1693 |
| Thermal | 3.53E-18 | 2.078 | 2518 | 2.78E-11 | 0.059 | 4496 |

## Acknowledgments

X. Li thanks Prof. Keli Han for his continuous encouragement and inspiring advices and Dr. Alan Heays for valuable discussions. The authors thank Drs. Erik Deul and Mark Somers for their assistance. This work is supported by the Dutch astrochemistry network (DAN) from the Netherlands Organization for Scientific Research (NWO) under grant 648.000.002. Astrochemistry in Leiden is also supported by the Netherlands Research School for Astronomy (NOVA), by a NWO Spinoza grant, and by the European Community's Seventh Framework Program FP7/2007-2013 under grant agreement 238258 (LASSIE).

# References

(1)     Garton, D. J.; Minton, T. K.; Maiti, B.; Troya, D.; Schatz, G. C. A Crossed Molecular Beams Study of the O($^3$P) + H$_2$ Reaction: Comparison of Excitation Function with Accurate Quantum Reactive Scattering Calculations. *J. Chem. Phys.* **2003**, *118*, 1585–1588.

(2)     Baulch, D. L.; Cobos, C. J.; Cox, R. A.; Esser, C.; Frank, P.; Just, T.; Kerr, J. A.; Pilling, M. J.; Troe, J.; Walker, R. W. Evaluated Kinetic Data for Combustion Modelling. *J. Phys. Chem. Ref. Data* **1992**, *21*, 411–734.

(3)     Graff, M. M.; Dalgarno, A. Oxygen Chemistry of Shocked Interstellar Clouds. II-Effects of Nonthermal Internal Energy on Chemical Evolution. *Astrophys. J.* **1987**, *317*, 432–441.

(4)     Sternberg, A.; Dalgarno, A. Chemistry in Dense Photon-Dominated Regions. *Astrophys. J. Suppl.* **1995**, *99*, 565–607.

(5)     Agúndez, M.; Cernicharo, J.; Goicoechea, J. R. Formation of Simple Organic Molecules in Inner T Tauri Disks. *Astron. Astrophys.* **2008**, *483*, 831–837.

(6)     Zhai, H.; Zhang, P.; Zhou, P. Quantum Wave Packet Calculation of the O($^3$P) + H$_2$ Reaction on the New Potential Energy Surfaces for the Two Lowest States. *Comput. Theor. Chem.* **2012**, *986*, 25–29.

(7)     Han, B.; Zheng, Y. Nonadiabatic Quantum Dynamics in O($^3$P) + H$_2$ → OH + H: A Revisited Study. *J. Comput. Chem.* **2011**, *32*, 3520–3525.

(8)     Liu, S.; Shi, Y. Influence of Vibrational Excitation on Stereodynamics for O($^3$P) + D$_2$ → OD + D Reaction. *Chinese J. Chem. Phys.* **2010**, *23*, 649–654.

(9)     Li, B.; Han, K.-L. Mixed Quantum-Classical Study of Nonadiabatic Dynamics in the O($^3$P$_{2,1,0}$, $^1$D$_2$) + H$_2$ Reaction. *J. Phys. Chem. A* **2009**, *113*, 10189–10195.

(10)    Pettey, L. R.; Wyatt, R. E. Application of the Moving Boundary Truncation Method to Reactive Scattering: H + H$_2$, O + H$_2$, O + HD. *J. Phys. Chem. A* **2008**, *112*, 13335–13342.

(11)    Weck, P. F.; Balakrishnan, N.; Brandão; Rosa, C.; Wang, W. Dynamics of the O($^3$P) + H$_2$ Reaction at Low Temperatures: Comparison of Quasiclassical Trajectory with Quantum Scattering Calculations. *J. Chem. Phys.* **2006**, *124*, 74308–74308-8.

(12)    Wang, W.; Rosa, C.; Brandao, J. Theoretical Studies on the O ($^3$P) + H$_2$ → OH + H Reaction. *Chem. Phys. Lett.* **2006**, *418*, 250–254.

(13)  Garashchuk, S.; Rassolov, V. A.; Schatz, G. C. Semiclassical Nonadiabatic Dynamics Based on Quantum Trajectories for the O($^3$P, $^1$D) + H$_2$ System. *J. Chem. Phys.* **2006**, *124*, 244307–244307-8.

(14)  Atahan, S.; Klos, J.; Zuchowski, P. S.; Alexander, M. H. An Ab Initio Investigation of the O($^3$P)-H$_2$($^1\Sigma^+_g$) van der Waals Well. *Phys. Chem. Chem. Phys.* **2006**, *8*, 4420–4426.

(15)  Chu, T.-S.; Zhang, X.; Han, K.-L. A Quantum Wave-Packet Study of Intersystem Crossing Effects in the O($^3$P$_{2,1,0}$, $^1$D$_2$) + H$_2$ Reaction. *J. Chem. Phys.* **2005**, *122*, 214301–214301-6.

(16)  Braunstein, M.; Adler-Golden, S.; Maiti, B.; Schatz, G. C. Quantum and Classical Studies of the O($^3$P) + H$_2$ ($v$ = 0-3, $j$ = 0) → OH + H Reaction Using Benchmark Potential Surfaces. *J. Chem. Phys.* **2004**, *120*, 4316–4323.

(17)  Brandão, J.; Mogo, C.; Silva, B. C. Potential Energy Surface for H$_2$O($^3$A") from Accurate Ab Initio Data with Inclusion of Long-Range Interactions. *J. Chem. Phys.* **2004**, *121*, 8861–8868.

(18)  Balakrishnan, N. Quantum Calculations of the O($^3$P) + H$_2$ → OH + H Reaction. *J. Chem. Phys.* **2004**, *121*, 6346–6352.

(19)  Maiti, B.; Schatz, G. C. Theoretical Studies of Intersystem Crossing Effects in the O($^3$P, $^1$D) + H$_2$ Reaction. *J. Chem. Phys.* **2003**, *119*, 12360–12371.

(20)  Balakrishnan, N. Quantum Mechanical Investigation of the O + H$_2$ → OH + H Reaction. *J. Chem. Phys.* **2003**, *119*, 195–199.

(21)  Rogers, S.; Wang, D.; Kuppermann, A.; Walch, S. Chemically Accurate Ab Initio Potential Energy Surfaces for the Lowest $^3$A' and $^3$A" Electronically Adiabatic States of O($^3$P) + H$_2$. *J. Phys. Chem. A* **2000**, *104*, 2308–2325.

(22)  Hoffmann, M. R.; Schatz, G. C. Theoretical Studies of Intersystem Crossing Effects in the O + H$_2$ Reaction. *J. Chem. Phys.* **2000**, *113*, 9456–9465.

(23)  Varandas, A. J. C.; Voronin, A. I.; Riganelli, A.; Caridade, P. J. S. B. Cross Sections and Rate Constants for the O($^1$D) + H$_2$ Reaction Using a Single-Valued Energy-Switching Potential Energy Surface. *Chem. Phys. Lett.* **1997**, *278*, 325–332.

(24)  Chatfield, D. C.; Friedman, R. S.; Lynch, G. C.; Truhlar, D. G.; Schwenke, D. W. The Nature and Role of Quantized Transition States in the Accurate Quantum Dynamics of the Reaction O + H$_2$ → OH + H. *J. Chem. Phys.* **1993**, *98*, 342–362.

(25) Joseph, T.; Truhlar, D. G.; Garrett, B. C. Improved Potential Energy Surfaces for the Reaction $O(^3P) + H_2 \rightarrow OH + H$. *J. Chem. Phys.* **1988**, *88*, 6982–6990.

(26) Robie, D. C.; Arepalli, S.; Presser, N.; Kitsopoulos, T.; Gordon, R. J. Evidence for Tunneling in the Reaction $O(^3P) + HD$. *Chem. Phys. Lett.* **1987**, *134*, 579–582.

(27) Schatz, G. C. A Coupled States Distorted Wave Study of the $O(^3P) + H_2$ ($D_2$ HD, DH) Reaction. *J. Chem. Phys.* **1985**, *83*, 5677–5686.

(28) Bowman, J. M.; Wagner, A. F.; Walch, S. P.; Thom. H. Dunning, J. Reaction Dynamics for $O(^3P) + H_2$ and $D_2$. IV. Reduced Dimensionality Quantum and Quasiclassical Rate Constants with an Adiabatic Incorporation of the Bending Motion. *J. Chem. Phys.* **1984**, *81*, 1739–1752.

(29) Schatz, G. C.; Wagner, A. F.; Walch, S. P.; Bowman, J. M. A Comparative Study of the Reaction Dynamics of Several Potential Energy Surfaces of $O(^3P)$ $+ H_2 \rightarrow OH + H$. I. *J. Chem. Phys.* **1981**, *74*, 4984–4996.

(30) Walch, S. P.; Wagner, A. F.; Thom. H. Dunning, J.; Schatz, G. C. Theoretical Studies of the $O + H_2$ Reaction. *J. Chem. Phys.* **1980**, *72*, 2894–2896.

(31) Howard, R. E.; McLean, A. D.; W. A. Lester, J. Extended Basis First-Order CI Study of the $^1A'$, $^3A''$, $^1A''$, and $\tilde{B}$ $^1A'$ Potential Energy Surfaces of the $O(^3P, ^1D)$ $+ H_2(^1\Sigma^+_g)$ Reaction. *J. Chem. Phys.* **1979**, *71*, 2412–2420.

(32) Light, G. C. The Effect of Vibrational Excitation on the Reaction of $O(^3P)$ with $H_2$ and the Distribution of Vibrational Energy in the Product OH. *J. Chem. Phys.* **1978**, *68*, 2831–2843.

(33) Garton, D. J.; Brunsvold, A. L.; Minton, T. K.; Troya, D.; Maiti, B.; Schatz, G. C. Experimental and Theoretical Investigations of the Inelastic and Reactive Scattering Dynamics of $O(^3P) + D_2$. *J. Phys. Chem. A* **2006**, *110*, 1327–1341.

(34) Westenberg, A. A.; Haas, N. de Atom―Molecule Kinetics Using ESR Detection. III. Results for $O + D_2 \rightarrow OD + D$ and Theoretical Comparison with $O + H_2 \rightarrow OH + H$. *J. Chem. Phys.* **1967**, *47*, 4241–4246.

(35) Baulch, D. L.; Cobos, C. J.; Cox, R. A.; Frank, P.; Hayman, G.; Just, T.; Kerr, J. A.; Murrells, T.; Pilling, M. J.; Troe, J. Evaluated Kinetic Data for Combustion Modelling. Supplement 1. *J. Phys. Chem. Ref. Data* **1994**, *23*, 847–1034.

(36) Sultanov, R. A.; Balakrishnan, N. Isotope Branching and Tunneling in $O(^3P) +$ $HD \rightarrow OH + D$; $OD + H$ Reactions. *J. Chem. Phys.* **2004**, *121*, 11038–11044.

(37)  Johnson, B. R.; Winter, N. W. Classical Trajectory Study of the Effect of Vibrational Energy on the Reaction of Molecular Hydrogen with Atomic Oxygen. *J. Chem. Phys.* **1977**, *66*, 4116–4120.

(38)  Reynard, L. M.; Donaldson, D. J. OH Production from the Reaction of Vibrationally Excited $H_2$ in the Mesosphere. *Geophys. Res. Lett.* **2001**, *28*, 2157–2160.

(39)  Wei, Q.; Li, X.; Li, T. Theoretical Study of Stereodynamics for the $O(^3P) + H_2$ ($v = 0$-2, $j=0$) → OH + H Reaction. *Chem. Phys.* **2010**, *368*, 58–61.

(40)  Chu, T. S.; Zhang, Y.; Han, K. L. The Time-Dependent Quantum Wave Packet Approach to the Electronically Nonadiabatic Processes in Chemical Reactions. *Int. Rev. Phys.Chem.* **2006**, *25*, 201–235.

(41)  Atahan, S.; Alexander, M. H.; Rackham, E. J. Cross Sections and Thermal Rate Constants for the Isotope Exchange Reaction: $D(^2S) + OH(^2\Pi) \rightarrow OD(^2\Pi) + H(^2S)$. *J. Chem. Phys.* **2005**, *123*, 204306–204306-10.

(42)  Howard, M. J.; Smith, I. W. M. Direct Rate Measurements on the Reaction D + OH → OD + H from 300 to 515 K. *J. Chem. Soc. Faraday Trans. 2 Mol. Chem. Phys.* **1982**, *78*, 1403–1412.

(43)  Dunne, L. J.; Murrell, J. N. Quasi-Classical Dynamics on the Ground State Surface of $H_2O$. *Mol. Phys.* **1983**, *50*, 635–644.

(44)  Margitan, J. J.; Kaufman, F.; Anderson, J. G. Kinetics of the Reaction OH + D → OD + H. *Chem. Phys. Lett.* **1975**, *34*, 485–489.

(45)  Radenovic, D. C.; van Roij, A. J. A.; Chestakov, D. A.; Eppink, A. T. J. B.; ter Meulen, J. J.; Parker, D. H.; van der Loo, M. P. J.; Groenenboom, G. C.; Greenslade, M. E.; Lester, M. I. Photodissociation of the OD Radical at 226 and 243 nm. *J. Chem. Phys.* **2003**, *119*, 9341–9343.

(46)  NIST Chemical Kinetics Database NIST 2013. *http://kinetics.nist.gov/kinetics/index.jsp*.

(47)  Cohen, N.; Westberg, K. R. Chemical Kinetic Data Sheets for High-Temperature Chemical Reactions. *J. Phys. Chem. Ref. Data* **1983**, *12*, 531–590.

(48)  Tappe, A.; Lada, C. J.; Black, J. H.; Muench, A. A. Discovery of Superthermal Hydroxyl (OH) in the HH 211 Outflow. *Astrophys. J. Lett.* **2008**, *680*, L117–L120.

(49)     Mandell, A. M.; Bast, J.; Dishoeck, E. F. van; Blake, G. A.; Salyk, C.; Mumma, M. J.; Villanueva, G. First Detection of Near-Infrared Line Emission from Organics in Young Circumstellar Disks. *Astrophys. J.* **2012**, *747*, 92(14pp).

(50)     Salyk, C.; Pontoppidan, K. M.; Blake, G. A.; Lahuis, F.; Dishoeck, E. F. van; II, N. J. E. $H_2O$ and OH Gas in the Terrestrial Planet-Forming Zones of Protoplanetary Disks. *Astrophys. J. Lett.* **2008**, *676*, L49–L52.

(51)     Pontoppidan, K. M.; Salyk, C.; Blake, G. A.; Meijerink, R.; Carr, J. S.; Najita, J. A Spitzer Survey of Mid-Infrared Molecular Emission from Protoplanetary Disks. I. Detection Rates. *Astrophys. J.* **2010**, *720*, 887–903.

(52)     Maillard, J. P.; Chauville, J.; Mantz, A. W. High-Resolution Emission Spectrum of OH in an Oxyacetylene Flame from 3.7 to 0.9 mm. *J. Mol. Spectrosc.* **1976**, *63*, 120–141.

(53)     W. L. Hase; R. J. Duchovic; X. Hu; A. Komornicki; K. F. Lim; D.-H. Lu; Peslherbe, G. H.; K. N. Swamy; S. R. V. Linde; A. Varandas; et al. VENUS96: A General Chemical Dynamics Computer Program . *Quantum Chem. Progr. Exch.* **1996**, *16,* 671.

(54)     Karplus, M.; Porter, R. N.; Sharma, R. D. Exchange Reactions with Activation Energy. I. Simple Barrier Potential for (H, $H_2$). *J. Chem. Phys.* **1965**, *43*, 3259–3287.

(55)     Ju, L. P.; Han, K. L.; Zhang, J. Z. H. Global Dynamics and Transition State Theories: Comparative Study of Reaction Rate Constants for Gas‐phase Chemical Reactions. *J. Comput. Chem.* **2009**, *30*, 305–316.

(56)     Truhlar, D. G.; Garrett, B. C. Variational Transition State Theory. *Annu. Rev. Phys. Chem.* **1984**, *35*, 159–189.

(57)     Fernandez-Ramos, A; Ellingson, B. A.; Garrett, B. C.; Truhlar, D. G. POLYRATE -9.7*, Univ. Minnesota, Minneap.* **2007**.

(58)     Tsang, W.; Hampson, R. F. Chemical Kinetic Data Base for Combustion Chemistry. Part I. Methane and Related Compounds. *J. Phys. Chem. Ref. Data* **1986**, *15*, 1087–1279.

**Figure Captions**

**Figure 1** Schematic energy-level diagram for the abstraction reactions H+OH $(\upsilon, j)$ → O ($^3$P) + H$_2$ at various rovibrational $(\upsilon, j)$ states of OH, in solid lines. Dotted lines: corrected energy levels which include the zero-point vibrational energy (ZPE). For the H + OH channel, this ZPE is taken into account in the QCT calculations.

**Figure 2** Relative populations of OH $(\upsilon, j)$ molecules as a function of $j$ for thermal excitation at various temperatures, for $\upsilon = 0$ and 1.

**Figure 3** Integral cross sections as functions of collision energy for the title reaction at different rovibrational levels $(\upsilon, j)$ of OH for the two PESs $^3$A' and $^3$A''.

**Figure 4** Typical reactive trajectories for the abstraction reaction H$_a$ + H$_b$O → O + H$_a$H$_b$ on the $^3$A' and $^3$A'' PESs, namely internuclear distances of H$_a$H$_b$, H$_b$O and H$_a$O as functions of propagation time.

**Figure 5** State-resolved rate constants as functions of inverse temperature for individual OH $(\upsilon, j)$ levels.

**Figure 6** Plot of rate constants for the title reaction with OH in the $(\upsilon = 0, j = 0)$ state, as functions of inverse temperature in the range of 200 – 2500 K, which were calculated from QCT and various TST methods. Rate constants at each electronic state were calculated by considering the results from two PESs, i.e., ($^3$A' + $^3$A'')/3.

**Figure 7** Comparison of the presently calculated thermal rate constants with that of Tsang *et al.*[47] The figure also illustrates the rotational and vibrational excitation effects on the rate constants. In the plot, x-axis is on a linear scale but the y-axis is on base 10 logarithmic scales. Rate constants for each electronic state were calculated by considering the results from two PESs, i.e., ($^3$A' + $^3$A'')/3.

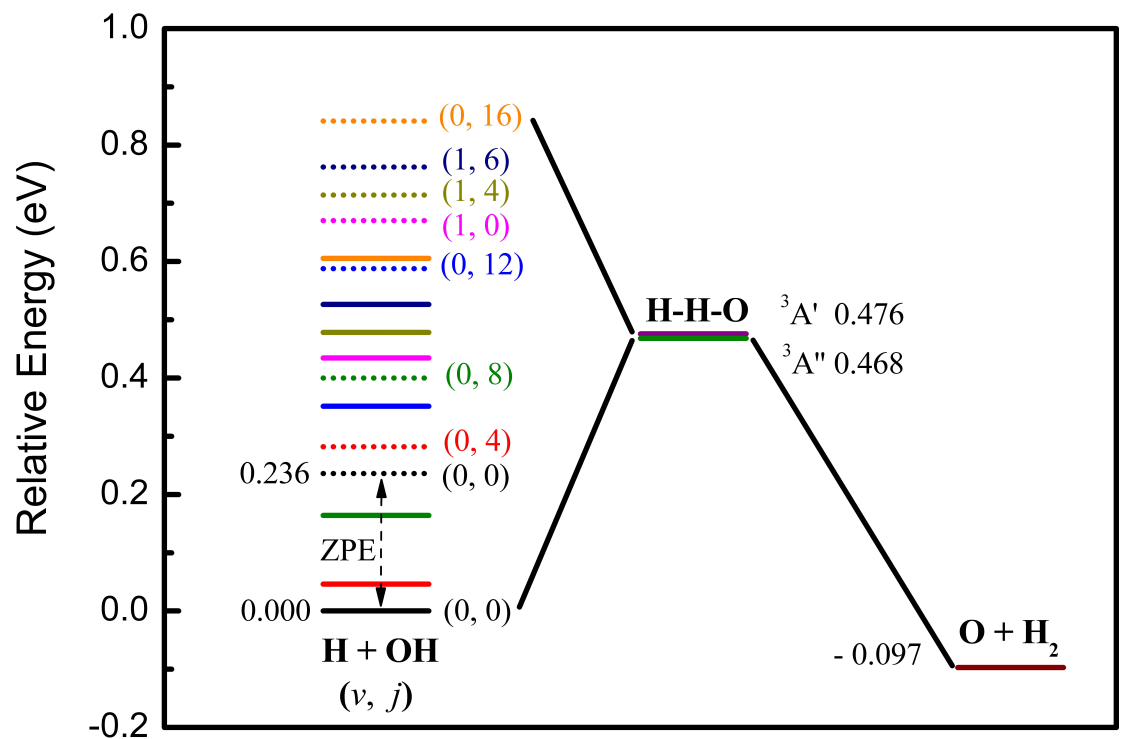

**Figure 1**

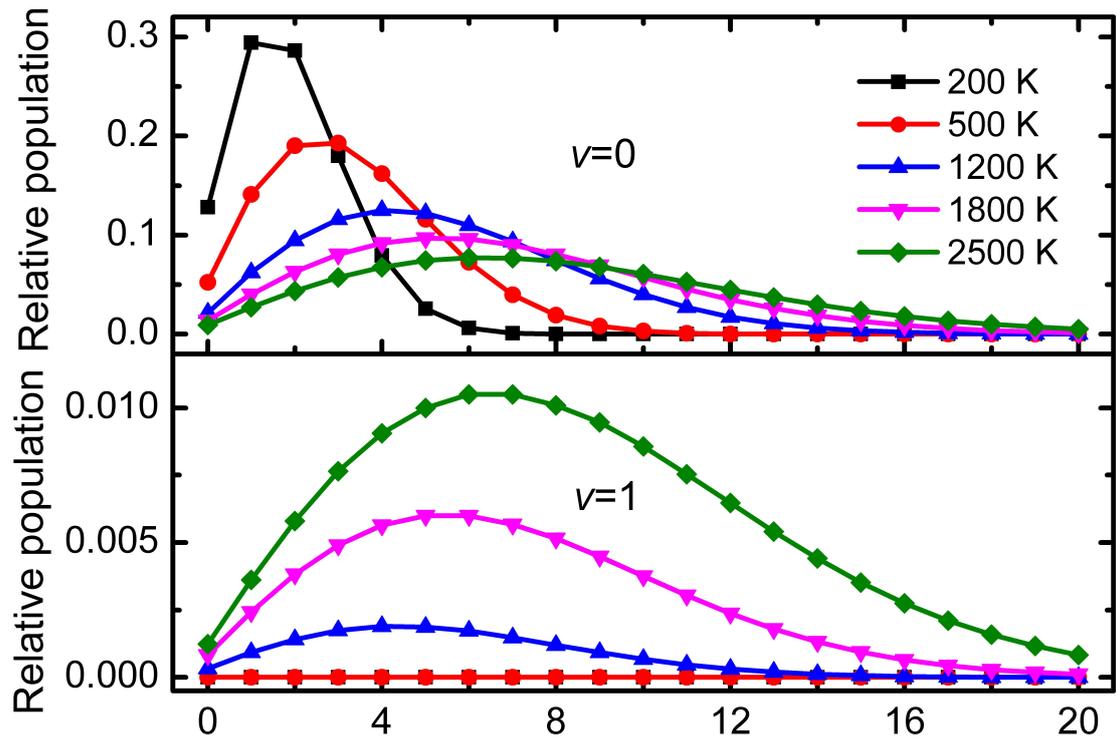

**Figure 2**

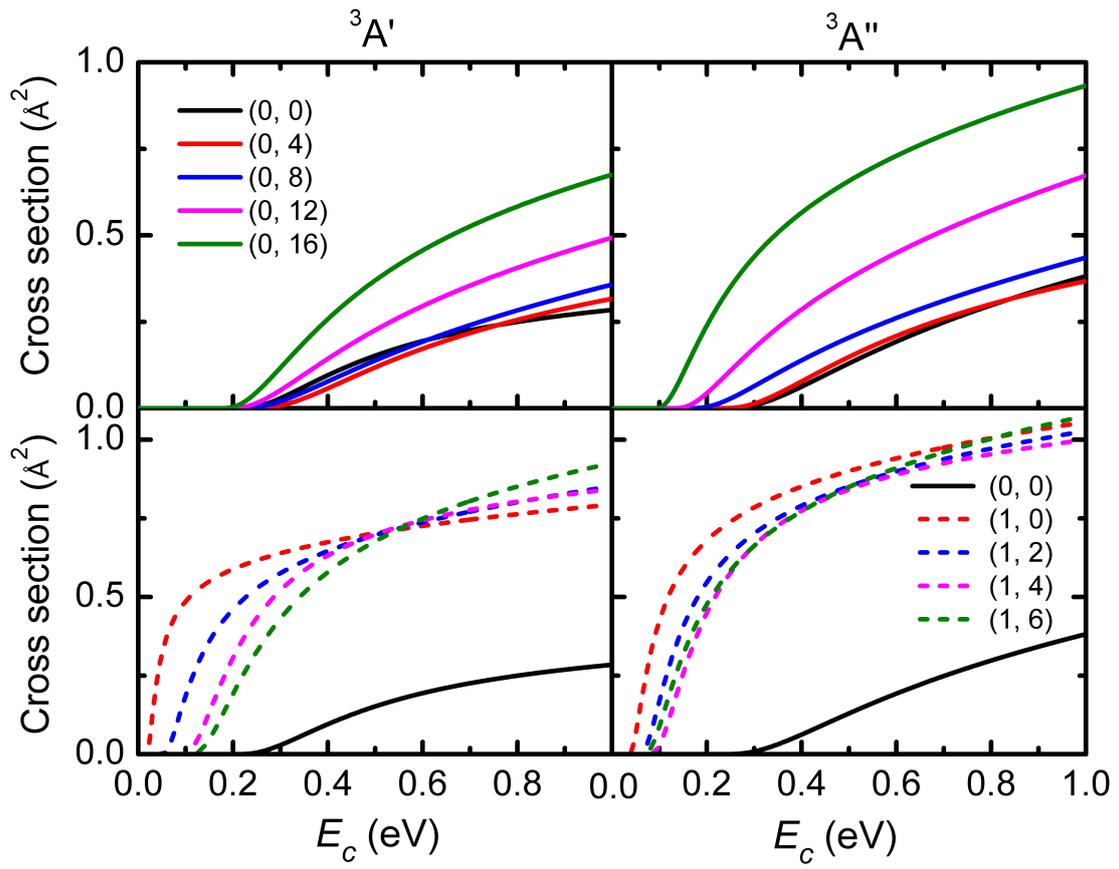

**Figure 3**

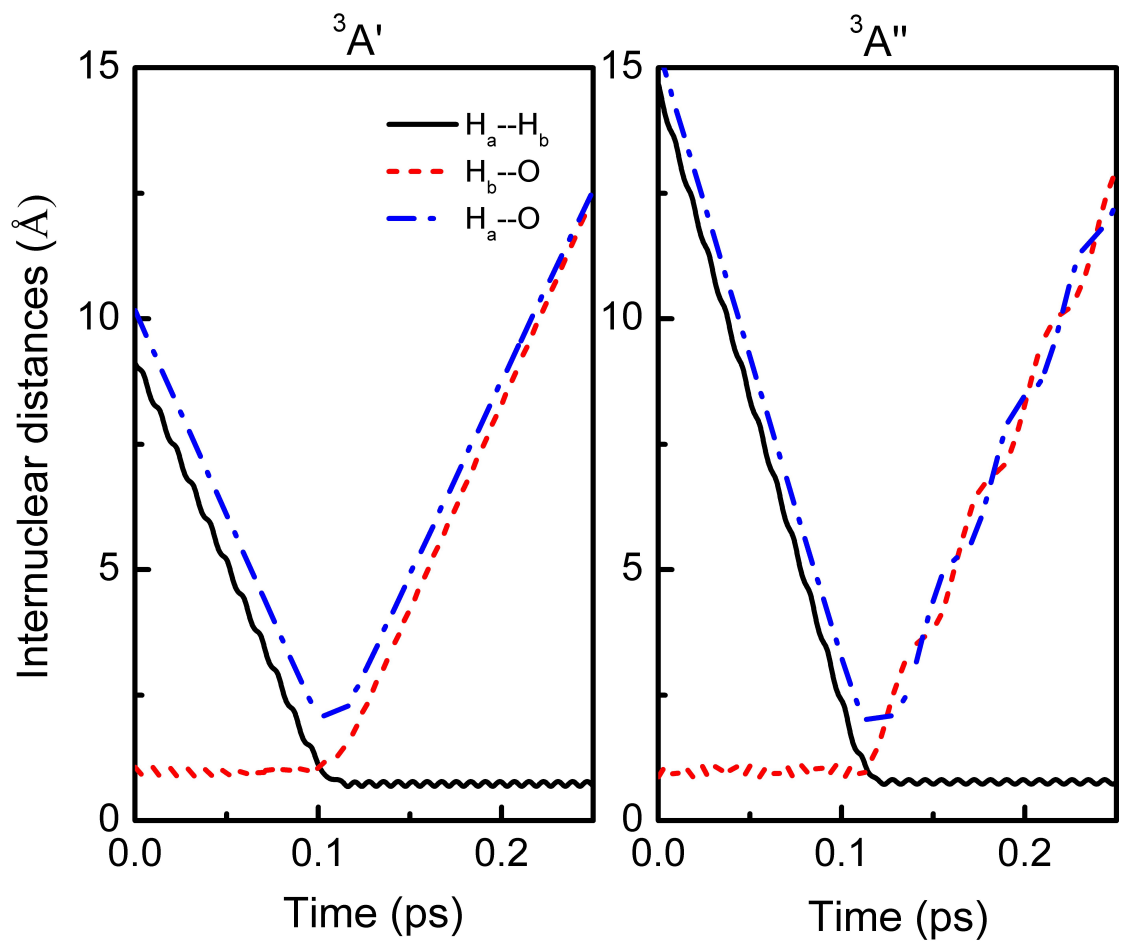

Figure 4

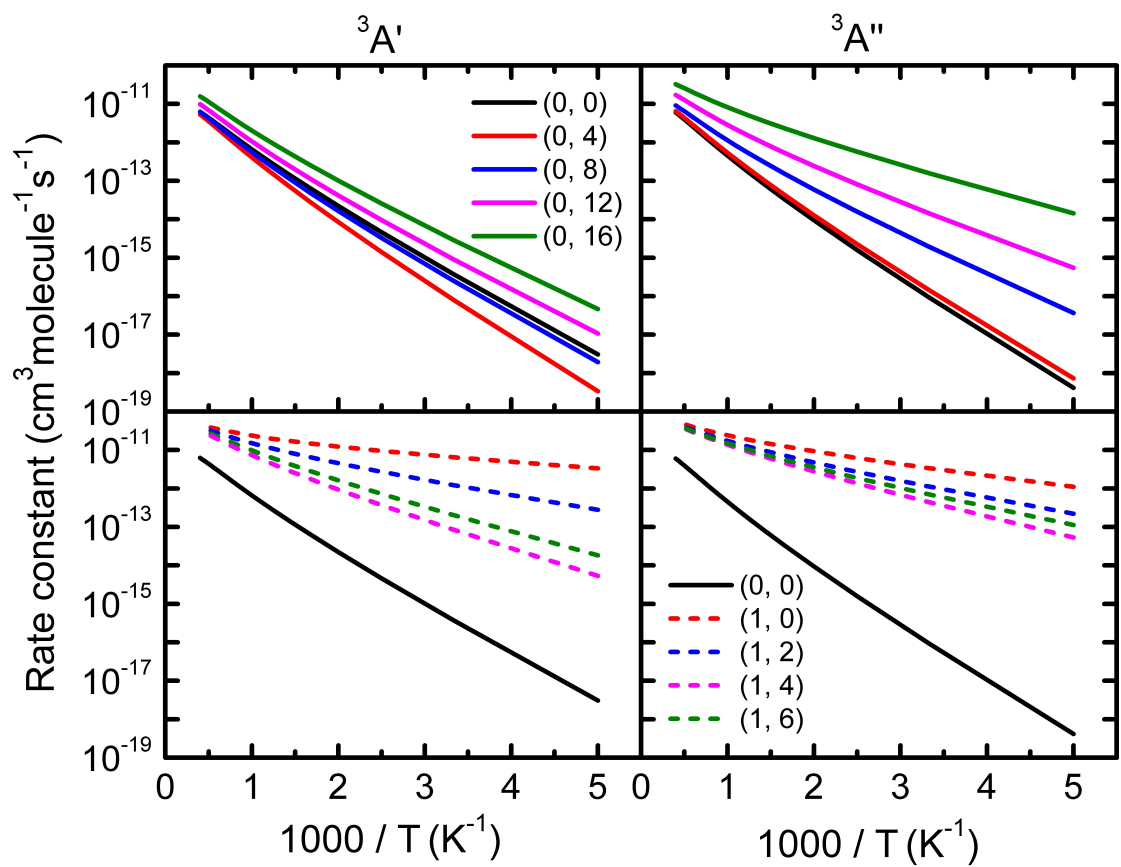

**Figure 5**

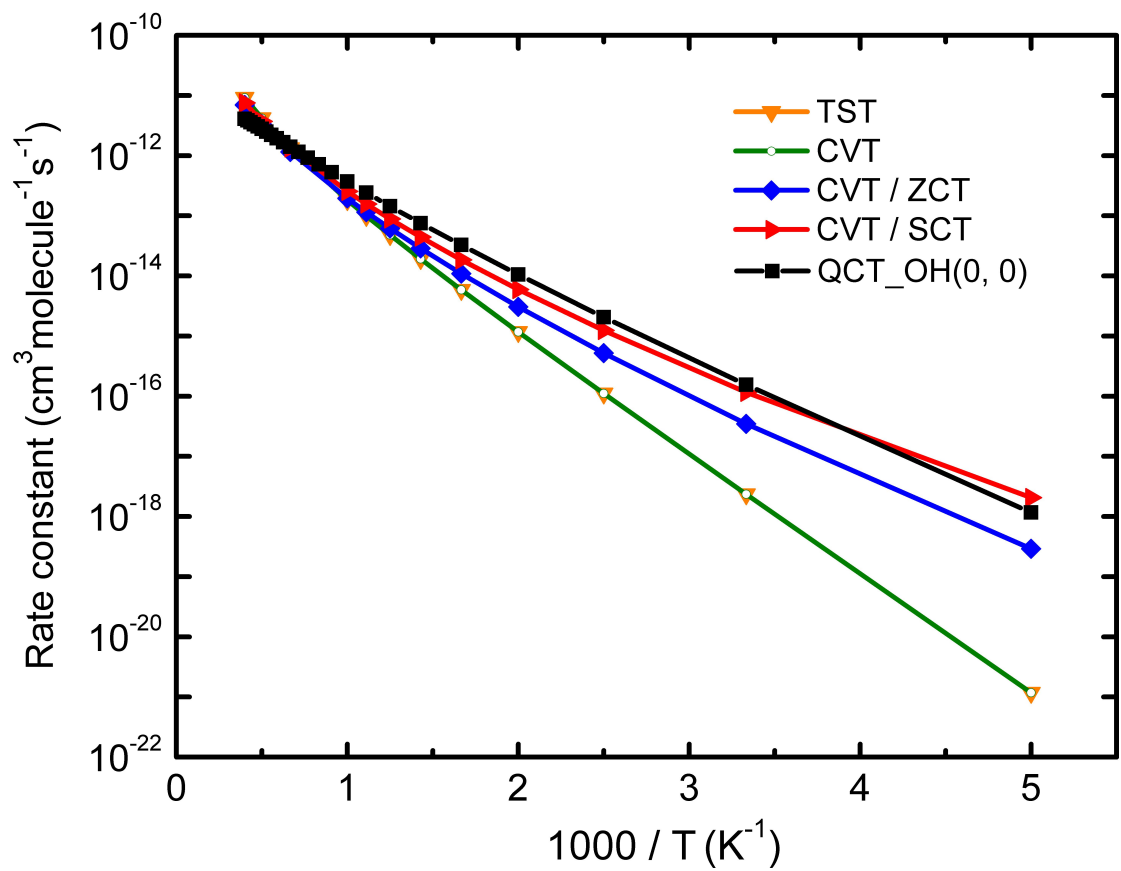

**Figure 6**

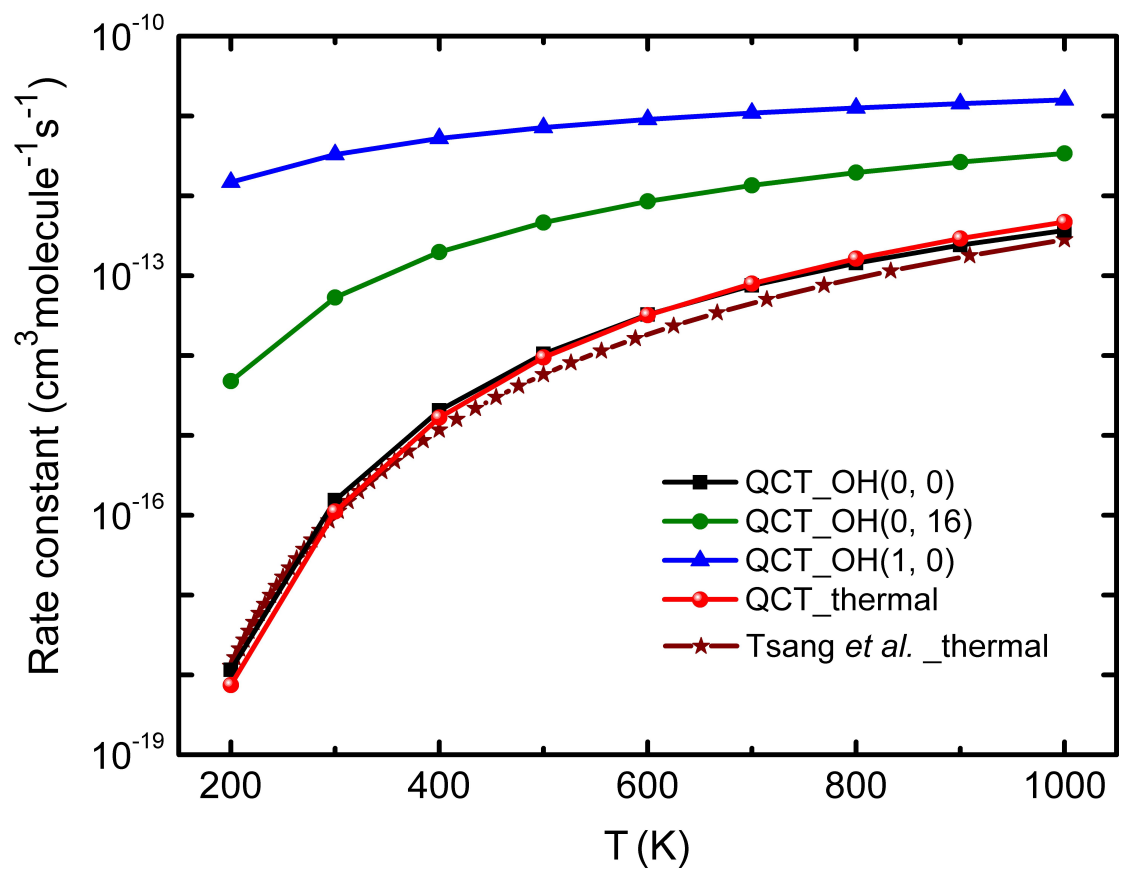

**Figure 7**

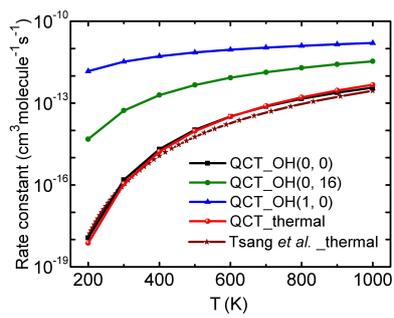

**Figure TOC**